\documentclass[preprint]{elsarticle}

\usepackage[T2A]{fontenc}
\usepackage[cp1251]{inputenc}
\usepackage{graphicx}
\usepackage{psfig}
\usepackage{amsmath}
\usepackage{caption2}
\usepackage{floatflt}
\usepackage{mathrsfs}

\journal{Superlattices and Microstructures}

\begin{document}
\begin{frontmatter}
\title{Internal Josephson phenomena in a coupled two-component Bose condensate}
\author[mymainaddress]{Nina~S.~Voronova\corref{mycorrespondingauthor}}
\cortext[mycorrespondingauthor]{Corresponding author}
\ead{nsvoronova@mephi.ru}
\address[mymainaddress]{National Research Nuclear University MEPhI (Moscow Engineering Physics Institute), Kashirskoe sh. 31, 115409 Moscow, Russia}
\author[mainaddress,secondaryaddress]{Yurii~E.~Lozovik}
\address[mainaddress]{Institute for Spectroscopy RAS, Fizicheskaya 5, 142190 Troitsk, Moscow, Russia}
\address[secondaryaddress]{Moscow Institute of Physics and Technology (State University), Institutskiy per. 9, 141700 Dolgoprudny, Moscow region, Russia}

\begin{abstract}
We discuss the coherent oscillations between two coupled quantum states of a Bose-Einstein condensate in two-dimensional space at zero temperature. In the system we consider, weak interparticle repulsive interactions occur between the particles in state one only, while the state two particles remain non-interacting. Analytical as well as numerical solution of the coupled Gross-Pitaevskii and Schr\"{o}dinger equations reveals various regimes of oscillational dynamics for the relative quantum phase and population imbalance between the two subsystems of the condensate. We show that, depending on the energy detuning between the two states, the system can exhibit modified harmonic and anharmonic Rabi oscillations or can transit to the regime analogous to internal a.c. Josephson effect. Morover, at a certain value of energy detuning, the internal oscillations are fully suppressed.
\end{abstract}


\end{frontmatter}

\section{Introduction}

Following the original discovery by Josephson in 1962 \cite{josephson}, the so-called external Josephson effect was extensively studied in superconducting \cite{barone,anderson2,backhaus}, superfluid helium \cite{anderson} and atomic bosonic weak links \cite{javanainen,smerzi,raghavan,zapata,giovanazzi,yu,albiez,bergeman,levy,diaz,nester,nester1} both theoretically  and experimentally. It has also been argued for a system of superfluid $^3$He-$A$ \cite{leggett,wheatley} that Josephson-like effects may occur for two condensates which are not spatially separated but occupy different quantum states. Later, experimental observation of internal Josephson effect in a two-state Bose-Einstein condensate (BEC) of $^{87}$Rb atoms was also reported \cite{matthews}. Theoretical research on the Josephson particle transfer between the ground and an excited state of an atomic Bose condensate is provided in Ref.\cite{yukalov}.

Recently, a new ground for the Josephson physics investigation appeared: microcavities \cite{verger,hennessy,gerace,QFL}. Great theoretical and experimental activity was aimed to study Josephson phenomena in exciton-polariton BECs coupled via a double-well potential \cite{wouters,shelykh,sarchi,read,lagoudakis,abbarchi}. Bosons participating in the phenomena are composite half-light half-matter quasiparticles interacting via their excitonic component \cite{MCpolaritons}, while the photon component allows them to condense at high critical temperatures \cite{kasprzak}. Despite the fact that different dynamical regimes which result from the interplay between the tunneling coupling strength of the two condensates and the interparticle interactions depend on the specific system under consideration \cite{wouters,lagoudakis,abbarchi}, the oscillations discussed in context of polaritons always imply that the two polaritonic BECs are spatially separated by a tunnel barrier. However, since polaritons itself are composite and result from strong coupling between the photon and quantum well exciton modes inside a planar optical cavity, they may be considered as a mixture of the two Bose-condensed components performing mutual transformations. The internal dynamics of an exciton-polariton gas was kept behind the scenes until the year of 2014, when several experimental \cite{laussy,sanvitto} and theoretical \cite{glazov,liew} works on polariton Rabi oscillations clearly demonstrated that there is a rich variety of dynamical behaviors in the system of exciton-polaritons still to be explored.

The present paper is dedicated to the study of various regimes of internal oscillations in the system of coupled linear (non-interacting) and non-linear (interacting) BECs occupying the same two-dimensional space, depending on the energy detuning between the two states. The theoretical problem in consideration is general, although we keep the polariton background in mind for the sake of realistic simulation parameters, which we extract from the recent microcavity experiments. Whereas for a pure quantum two-level system with no interactions the dynamics is described by single-particle harmonic Rabi oscillations, we show that in a more sophisticated system such as BEC with interactions and non-zero energy detuning, various scenarios of \textit{collective} non-linear oscillations become possible, including harmonic and anharmonic modifications of Rabi oscillations, the analog of internal a.c. Josephson effect, and full suppression of oscillations.

The paper is organized as follows. In Section~\ref{sec2} we introduce all notations and the theoretical model, and analyze the time-dependent coupled equations describing temporal dynamics of the system. In Section~\ref{sec3} we solve the evolution equations, both analytically and numerically, and discuss various dynamical regimes of the system which can be realized, depending on the parameters. Section~\ref{conclusion} summarizes our results.

\section{Two-mode Gross-Pitaevskii model}\label{sec2}

Within the mean-field approximation, ground-state unit area energy functional for weakly nonhomogeneous two-component Bose condensate with weak repulsive interactions in one component can be introduced as \cite{j.nanophot}:
\begin{multline}\label{e_func}
\mathscr{E}[\psi_1,\psi_1^*,\psi_2,\psi_2^*] = -\frac{\hbar^2}{2m_1}\,\psi_1^*\nabla\psi + E_1^0|\psi_1|^2+\frac{\bar{g}}{2}|\psi_1|^4 \\
-\frac{\hbar^2}{2m_2}\,\psi_2^*\nabla\psi_2 + E_2^0|\psi_2|^2 +\frac{\hbar\Omega_R}{2}(\psi_1^*\psi_2+\psi_2^*\psi_1),
\end{multline}
with $\psi_{1,2}$ the two macroscopic wave functions of the condensate components, $E_{1,2}^0$ the bottoms of energy dispersions and $m_{1,2}$ the effective masses of the two types of particles. $\bar{g}>0$ is the constant of repulsive interaction in the 1st (interacting) component. Particle transfer between the subsystems is described by the coupling term $\sim\hbar\Omega_R/2$, where $\hbar\Omega_R$ is the Rabi splitting energy. 

Using variational principle $i\hbar\partial_t\psi_{1,2}=\delta\mathscr{E}/\delta\psi_{1,2}^*$, one gets a set of two coupled differential equations of the Gross-Pitaevskii type,
\begin{equation}\label{GPE_1}
i\hbar\partial_t\psi_1 = \Bigl[E_1^0-\frac{\hbar^2\nabla^2}{2m_1}+\bar{g}|\psi_1|^2\Bigr]\psi_1 + \frac{\hbar\Omega_R}{2}\,\psi_2,
\end{equation}
\begin{equation}\label{GPE_2}
i\hbar\partial_t\psi_2=\Bigl[E_2^0-\frac{\hbar^2\nabla^2}{2m_2}\Bigr]\psi_2 + \frac{\hbar\Omega_R}{2}\,\psi_1.
\end{equation}

Since we are interested in the temporal evolution of the system and consider no trapping potential, we restrict our consideration to the homogeneous case when the wave functions profiles are spatially uniform. Therefore we omit all the spatial derivatives in Eqs.~(\ref{GPE_1}) and (\ref{GPE_2}). Further in this paper, we rescale lengths and energies in terms of harmonic-oscillator units $\sqrt{\hbar/m_1\Omega_R}$ and $\hbar\Omega_R$, respectively, time as $t\Omega_R \rightarrow t$ and the wavefunctions as $\psi_{1,2}/\sqrt{\hbar/m_1\Omega_R}\rightarrow\psi_{1,2}$. 
After the Madelung transformation $\psi_{1,2}(t)=\sqrt{n_{1,2}(t)}\,e^{iS_{1,2}(t)}$, we introduce a new set of variables: population imbalance between the two subsystems $\rho(t) = n_1(t)-n_2(t)$ and the relative quantum phase $S(t)=S_1(t)-S_2(t)$. These variables $\rho$ and $S$ obey the coupled evolution equations
\begin{equation}\label{rho}
\dot{\rho} = -\sqrt{n^2-\rho^2}\,\sin S,
\end{equation}
\begin{equation}\label{s}
\dot{S} =  -\Delta E -\Lambda\frac{\rho}{n} + \frac{\rho}{\sqrt{n^2-\rho^2}}\,\cos S,
\end{equation}
where $n=n_1(t)+n_2(t)$ is total number of particles in the condensate. The dimensionless effective detuning $\Delta E=\delta + gn/2$ (where $\delta=\epsilon^0_1-\epsilon^0_2$, $\epsilon^0_{1,2}=E^0_{1,2}/\hbar\Omega_R$) and the dimensionless blueshift value $\Lambda = gn/2$ ($g =\bar{g} m_1/\hbar^2$) are the parameters which determine different regimes of the system behavior. It is worth noting that for a closed conservative system that we consider the equations (\ref{rho}), (\ref{s}) have Hamiltonian form for canonically conjugate variables: $\dot{\rho} = \partial H/\partial S$, $\dot{S} = -\partial H/\partial \rho$. The conserved energy is
\begin{equation}\label{hamilt}
H(S,\rho) = \Delta E\rho + \Lambda\,\frac{\rho^2}{2n} + \sqrt{n^2-\rho^2}\,\cos S,
\end{equation}
where total population $n$ remains constant. The Hamiltonian (\ref{hamilt}) is analogous to that of classical nonrigid pendulum with length dependent on its angular momentum $\rho$ (although, due to the choice of sign in front of the coupling term $\sim\hbar\Omega_R/2$ in (\ref{e_func}), the pendulum's ``restoring force'' is driving the tilt angle $S$ to $\pi$ instead of zero). 

\section{Results and discussion}\label{sec3}
\subsection{Symmetric case $\Delta E=0$}\label{sec3_1}

For the starting analysis we consider the system with no effective detuning: $\Delta E=0$ (\textit{i.e.} bare detuning $\delta$ compensates the blueshift $gn/2$). In this case, Eqs. (\ref{rho}) and (\ref{s}) may be transformed to the following:
\begin{equation}\label{rho_harmonic}
\ddot{\rho}+\rho\left(1+\Lambda\sqrt{1-\left(\frac{\rho}{n}\right)^2}\cos S\right)=0,
\end{equation}
\begin{equation}\label{S_harmonic}
\ddot{S}+\frac{1}{2}\frac{n^2+\rho^2}{n^2-\rho^2}\,\sin(2S)-\frac{\Lambda}{\sqrt{1- \left(\frac{\rho}{n}\right)^2}}\,\sin S=0.
\end{equation}

If interactions are negligible ($\Lambda\rightarrow 0$), Eq. (\ref{rho_harmonic}) reduces to $\ddot{\rho}+\rho=0$ and describes a harmonic modification of Rabi oscillations between the ``state 1'' and ``state 2'' with natural frequency $\Omega_R$ (which corresponds to $\omega=1$ in unscaled units) and time-average value $\langle \rho\rangle=0$. Eq. (\ref{S_harmonic}) in general case is referred to as the Hill equation and describes a so-called parametric oscillator. The explicit solutions for relative phase and population imbalance for $\Lambda=0$ read
\begin{equation}\label{rho-solution}
\rho(t) = \mp\sqrt{n^2-H_0^2}\sin (t\mp t_0),
\end{equation}
\begin{equation}\label{s-solution}
\cos S(t) = \frac{H_0}{\sqrt{n^2\cos^2(t\mp t_0)+H_0^2\sin^2(t\mp t_0)}},
\end{equation}
where $H_0 = H[S(0),\rho(0)] = \sqrt{n^2-\rho^2(0)}\cos S(0)$ and initial phase of oscillations $t_0=\arcsin[\rho(0)/\sqrt{n^2-H_0^2}]$. The upper sign (`$-$') corresponds to the case when $\pi/2<S(0)<\pi$, the lower sign (`$+$') corresponds to the case when $\pi<S(0)<3\pi/2$ (for $\pi$-phase modes with $\langle S\rangle=\pi$).

The shape of $S(t)$ temporal profile is strongly dependent on the initial conditions. In case of small-amplitude oscillations around $\pi$, $S(t)$ is quasiperiodic with a modulated ``period'' $T(t) = 2\pi/\sqrt{(n^2+\rho^2(t))/(n^2-\rho^2(t))}$ ($0<T\leq2\pi$). Furthermore, in the limit $\rho\ll n$, Eq. (\ref{S_harmonic}) reduces to $\ddot{S}+S=\pi$, and hence the relative phase oscillates with the same frequency $\Omega_R$ as does the population imbalance $\rho(t)$. If $|S(0)-\pi|$ is comparable to $\pi/2$ or $\rho(0)\rightarrow n$, $S(t)$ becomes strongly unharmonic while $\rho(t)$ oscillates with large amplitude $\sqrt{n^2\sin^2 S(0)+\rho^2(0)\cos^2 S(0)}$. 

\begin{figure}[t]
\renewcommand{\captionlabeldelim}{.}
{\center
\includegraphics[width=0.6\textwidth]{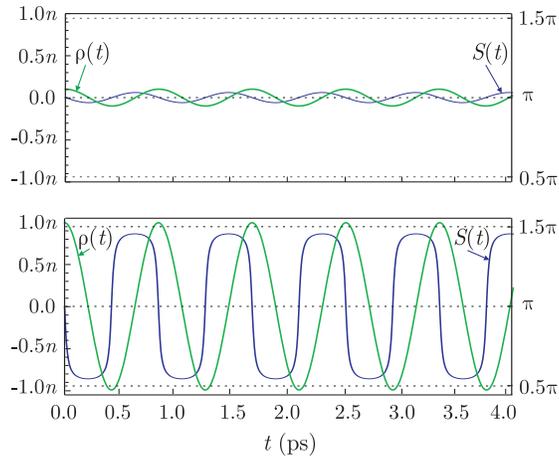}
\caption{\small (Color online) Population imbalance $\rho=n_1-n_2$ and relative quantum phase $S=S_1-S_2$ as functions of time. The initial population imbalance $\rho(0)$ takes the values $0.1n$ (top) and $0.99n$ (bottom). Numerical results of the simulations of Eqs. (\ref{rho}) and (\ref{s}) coincide with the analytical solutions (\ref{rho-solution}) and (\ref{s-solution}) for the values of the parameters used: $\hbar\Omega_R=5$~meV, $\bar{g}=0.015$~meV$\cdot\mu$m$^2$.}\label{fig01}
}
\end{figure}

Fig.~\ref{fig01} shows population imbalance and relative phase against time for the case of small-amplitude oscillations ($\rho(0)=0.1n$) and for large-amplitude oscillations ($\rho(0)=0.99n$). Speaking of polaritons, these two different initial states may be prepared with the help of different pumping schemes. Upper panel of Fig.~\ref{fig01} assumes that an almost equal population of coherent excitons and photons is taken as initial condition, which can be achieved with non-resonantly excited polariton condensate. Resonant excitation would imply that polaritons are created through their photonic component, then the initial conditions will correspond to the situation shown in the lower panel of Fig.~\ref{fig01}. It is also worth noting that the lower polariton state corresponds to the relative phase between the photon and exciton subsystems equal to $\pi$ (due to the positive sign chosen in Eqs.~(\ref{GPE_1}), (\ref{GPE_2}) in front of the coupling term $\hbar\Omega_R/2$), therefore here and below we will always imply the initial phase difference $S(0)=\pi$ unless stated otherwise. If there is no initial population imbalance $\rho(0)=0$, the system will stay in the pure Rabi regime which consists of density oscillations (with the imbalance always zero) without oscillations of the relative phase.

In the context of exciton-polaritons, the unscaled interaction constant $g$ is of the order of $10^{-3}$ (estimated from $\bar{g}=0.015$~meV$\cdot\mu$m$^2$ \cite{kasprzak}). Thus, the type of behavior described above (for $\Lambda\rightarrow 0$) can change only for large values of the total density $n$ of the polariton condensate. Numerical solution of Eqs. (\ref{rho_harmonic}) and (\ref{S_harmonic}) which takes into account interactions ($\Lambda\neq 0$) starts to noticeably differ from the analytical solution given by (\ref{rho-solution}) and (\ref{s-solution}) only when the parameter $\Lambda$ becomes of the order of $10^{-1}$ and larger.
For small-amplitude oscillations, this difference appears as a shift of the oscillation frequency $\omega=\Omega_R\sqrt{1 + \Lambda}$. For large-amplitude oscillations, when one considers large values of $\rho$ comparable to $n$, oscillations frequency becomes dependent on the initial values $\rho(0)$ and $S(0)$. The regimes of oscillations at different values of $\Lambda$ are not shown here as they are completely analogous to those appearing in the bosonic Josephson junction thoroughly discussed in Ref.~\cite{raghavan}. For $g\sim10^{-3}$, however, reaching even $\Lambda\sim1$ would require condensate densities as large as $n\sim10^3$. N.B., for realistic polariton densities, the unscaled $n$ is of the order of unity (estimated from $10^{10}$~cm$^{-2}$ \cite{kasprzak}), hence a system with $n=$~const this regime is practically not realized, and the effect of interactions on the modified Rabi dynamics stays negligible.

\subsection{Asymmetric case $\Delta E\neq 0$}

Let us now focus at the main goal of the present research: the case of non-zero effective detuning, in which the system behavior is influenced by the additional term $(\delta+\Lambda)\rho$ in the Hamiltonian (\ref{hamilt}). From the mathematical point of view, the resulting dynamics should be governed by a competition of the terms $\Lambda\rho/n$ and $\Delta E$ in Eq. (\ref{s}). Physically, the detuning between the modes is assumed to be less than or comparable to the Rabi splitting energy $\hbar\Omega_R$, which in unscaled units used in this paper means that the absolute value of $\delta$ ranges from zero to a number of the order one. Therefore, since the maximum ratio $\rho/n$ equals 1, for almost all values of $\delta$ the term containing $\Delta E$ wins this competition. 

\begin{figure}[t]
\renewcommand{\captionlabeldelim}{.}
{\center
\includegraphics[width=\textwidth]{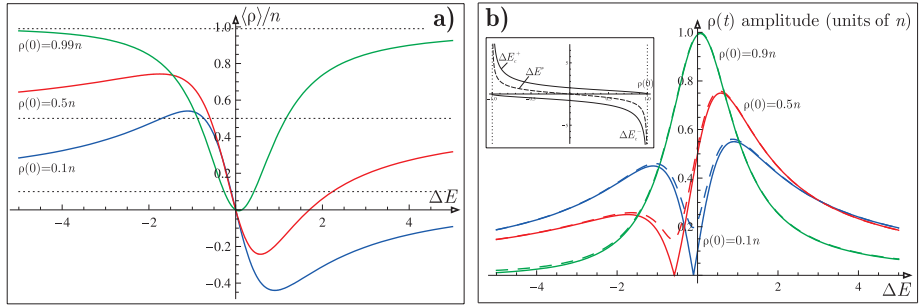}
\caption{\small (Color online) (a) Normalized time-average value of population imbalance $\langle\rho\rangle/n$ as a function of $\Delta E$ (given in the text by (\ref{avr})) for initial values $\rho(0)=0.1n$, $0.5n$, $0.99n$ as marked. Each curve has extrema at the points $\Delta E=\Delta E_c^{\pm}$ and tends to the corresponding $\rho(0)$ (dashed lines) as $\Delta E\rightarrow\pm\infty$. (b) Amplitude of population imbalance oscillations as a function of $\Delta E$ (analytical expression not shown) for initial values $\rho(0)=0.1n$, $0.5n$, $0.9n$ as marked. Solid lines correspond to $S(0)=\pi$, dashed lines correspond to $S(0)=\pi+0.2$. At $\Delta E=\Delta E_c^{\pm}$, the amplitude reaches its maxima. At $\Delta E=\Delta E^\star$, the amplitude has a minimum (zero, in case $S(0)=\pi$). Note that for negative initial values $\rho(0)<0$, both pictures will flip respectively to the $Y$-axis. The inset shows the values $\Delta E_c^{\pm}$ (solid lines) and $\Delta E^\star$ (dashed line) against $\rho(0)$.}
\label{avr_fig}
}
\end{figure}

Whereas for the case $\Delta E=0$ and small $\Lambda$ the time-average $\langle\rho\rangle$ was always zero, for $\Delta E\neq0$ the oscillations of $\rho(t)$ always average to a non-zero number dependent on $\Lambda$, $\Delta E$ and the initial conditions. For vanishing $\Lambda$,
\begin{equation}\label{avr}
\langle\rho\rangle\rightarrow\Delta E\,\frac{\Delta E\rho(0)+\sqrt{n^2-\rho^2(0)}\,\cos S(0)}{1+(\Delta E)^2}.
\end{equation}
Thus, for a given $\rho(0)$, the shift of $\langle\rho\rangle$ is determined by $\Delta E$. This is analogous to an external magnetic field applied to a mixture of spin-up and spin-down atoms with populations performing Josephson-like oscillations. Fig.~\ref{avr_fig}(a) shows $\langle\rho\rangle$ against $\Delta E$ for $\rho(0)=0.1n$, $0.5n$, $0.99n$.

\begin{figure}[t]
\renewcommand{\captionlabeldelim}{.}
{\center
\includegraphics[width=\textwidth]{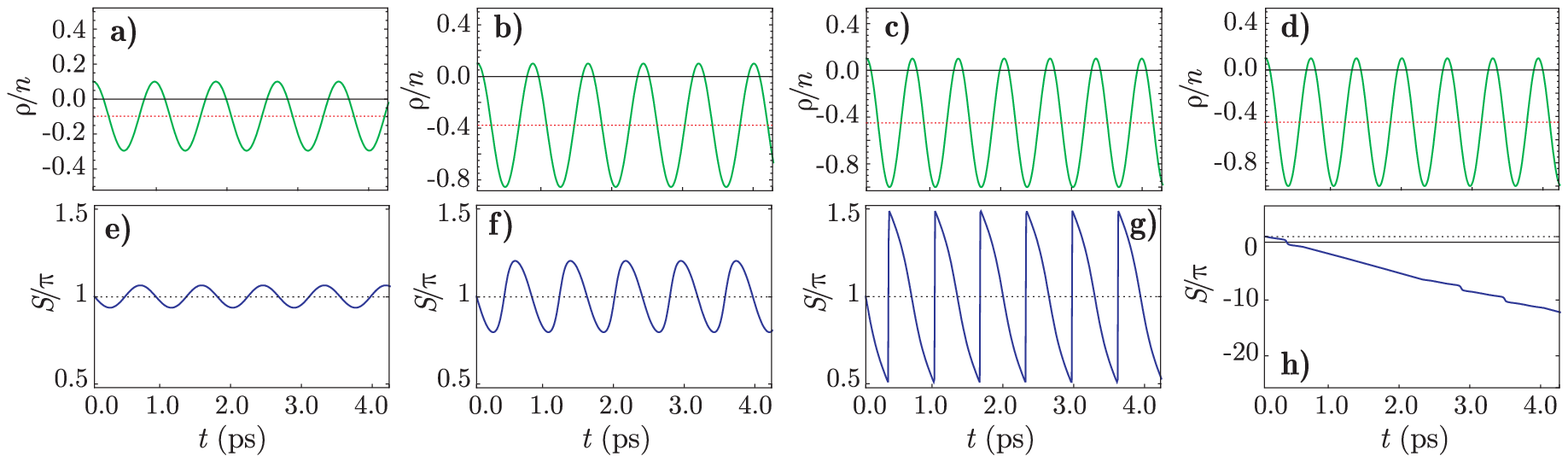}
\caption{\small (Color online)  Normalized population imbalance $\rho(t)/n$ and relative phase $S(t)$ (in units of $\pi$) as functions of time, with the initial conditions $\rho(0)=0.1n$, $S(0)=\pi$. Red/light dotted lines represent the normalized time-averaged $\langle\rho\rangle/n$ given by (\ref{avr}). Values of the effective detuning $\Delta E$: (a) and (e) $\Delta E=0.1\Delta E_c^{+}$, (b) and (f) $\Delta E=0.55\Delta E_c^{+}$, (c) and (g) $\Delta E=\Delta E_c^{+}$, (d) and (h) $\Delta E=1.006\Delta E_c^{+}$. Other physical values same as in Fig.~\ref{fig01}.}
\label{de}
}
\end{figure}

Frequency of oscillations in $\rho(t)$ for $\Lambda\rightarrow0$ is equal to $\Omega_R\sqrt{1+(\Delta E)^2}$. Here, the interplay between the Rabi and the ``a.c. Josephson'' dynamics becomes apparent: for vanishing values of $\delta$ and $\Lambda$, when $|\Delta E|\ll1$, we arrive at the Rabi regime discussed in the previous Subsection. In the case of larger detunings which result in $|\Delta E|$ comparable to or larger than 1, the oscillations between the two condensates are entirely analogous to the internal a.c. Josephson effect with the detuning playing a role of external voltage. For the intermediate values of $|\Delta E|$, there are several regimes possible, which differ in the behavior of the relative phase $S(t)$. These regimes of oscillations are shown in Fig.~\ref{de}.

For positive detunings and $\rho(0)>0$ (or, equivalently, for negative detunings and $\rho(0)<0$), as $|\Delta E|$ increases, harmonic Rabi oscillations in $\rho(t)$ shift from $\langle\rho\rangle=0$ according to (\ref{avr}) and grow in amplitude. This is shown in Fig.~\ref{de}(a)--(c). Oscillations of the relative phase $S(t)$ lose harmonicity and become of the shape of smoothed sawtooth with the amplitude growing up to $\pi/2$ when $\Delta E$ approaches its critical value $\Delta E_c^{+}$ (see Fig.~\ref{de}(e)--(g)). This critical value $\Delta E_c^{+}$ is dependent on the initial values $\rho(0)$ and $S(0)$. 
More precisely, for $\Lambda\rightarrow0$ we have
\begin{equation}\label{Ec-plus}
\Delta E_c^{+} = \frac{\rho(0) - \sqrt{\rho^2(0)\sin^2S(0)+n^2\cos^2S(0)}}{\sqrt{n^2-\rho^2(0)}\cos S(0)}.
\end{equation}
As follows from (\ref{Ec-plus}), for $\rho(0)\rightarrow 0$ the critical $\Delta E_c^{+}\rightarrow 1$ (which means positive detuning $\rightarrow\hbar\Omega_R$), and for $\rho\rightarrow n$ it tends to zero. 
The dependence (\ref{Ec-plus}) is plotted in the inset of Fig.~\ref{avr_fig}(b) for $S(0)=\pi$.

When $\Delta E$ exceeds $\Delta E_c^{+}$, the relative phase becomes running in time while $\langle\rho\rangle$ starts to shift in the opposite direction (see (\ref{avr}) and Fig.~\ref{avr_fig}(a)) and the amplitude of oscillations begins to decay (amplitude against $\Delta E$ is shown in Fig.~\ref{avr_fig}(b)). This latter regime of running phase is shown in Fig.~\ref{de}(d) and \ref{de}(h), and it is a clear manifestation of the a.c. Josephson effect, while the transitional regime of the sawtooth phase profile (Fig.~\ref{de}(c) and \ref{de}(g)) can be explained as ``beats'' between the Rabi and Josephson modes. If $\Delta E$ could be increased infinitely, $\langle\rho\rangle$ would tend to $\rho(0)$ which lies to the other side from zero compared to the initial shift of $\langle\rho\rangle$ and the amplitude of oscillations would tend to zero, since in the limit $\Delta E\rightarrow\infty$ the system becomes uncoupled.

For the case when $\Delta E$ and $\rho(0)$ have opposite signs, the dynamics possesses certain differences. A peculiar feature is that when $|\Delta E|$ starts increasing, amplitude of oscillations in $S(t)$ and $\rho(t)$ decreases. If $S(0)=\pi$, the amplitudes drop down to zero at $\Delta E=\Delta E^\star=(\rho(0)\cos S(0)+n\sin S(0)) /\sqrt{n^2-\rho^2(0)}$. The oscillations appear fully suppressed, and the values of population imbalance and the relative phase are fixed at $\rho=\rho(0)$ and $S=S(0)=\pi$. In case $S(0)\neq\pi$, the amplitudes reach local minima with values dependent on $S(0)$ and $\rho(0)$ (see Fig.~\ref{avr_fig}(b)). When $\Delta E>\Delta E^\star$, the amplitudes start to grow and the relative phase acquires the sawtooth-like temporal profile, analogous to the case discussed above. Upon reaching the critical value
\begin{equation}\label{ec-minus}
\Delta E_c^{-} = \frac{\rho(0) + \sqrt{\rho^2(0)\sin^2S(0)+n^2\cos^2S(0)}}{\sqrt{n^2-\rho^2(0)}\cos S(0)},
\end{equation}
the behavior is once more changed to the a.c. Josephson regime characterized by the running phase. Another difference with the previous case is that when $\rho(0)\rightarrow n$, the critical value $\Delta E_c^{-}$ goes to \begin{figure}[h]
\renewcommand{\captionlabeldelim}{.}
{\center
\includegraphics[width=0.8\textwidth]{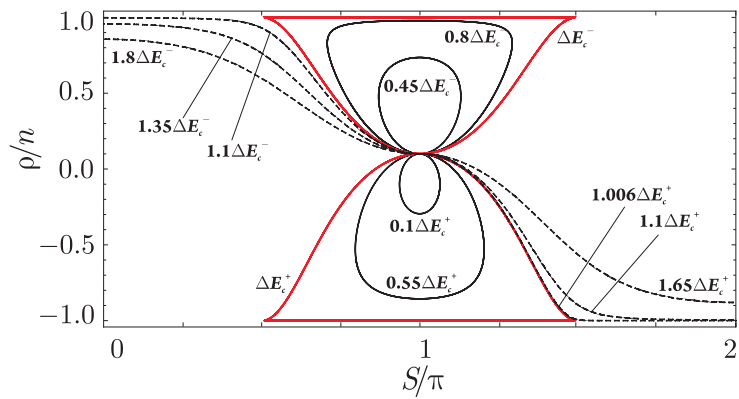}
\caption{\small (Color online) Phase-plane portrait of the conjugate variables $\rho$ (in units of $n$) and $S$ (in units of $\pi$) for $\Delta E\neq0$: all trajectories for $\rho(0)=0.1n$ and $\Delta E$ values as marked. Other physical values same as in Fig.~\ref{fig01}. See the text for more explanations.}
\label{phase-plane}
}
\end{figure}
infinity (see the inset of Fig.~\ref{avr_fig}(b)), which means for large $\rho(0)$, in the case $\rho(0)\Delta E<0$, the crossover to the Josephson regime doesn't occur, and $\langle\rho\rangle$ shifts in the direction of the initial value $\rho(0)$.

Phase-plane portrait $S(\rho)$ for the case $\Delta E\neq0$ is shown in Fig.~\ref{phase-plane}. All trajectories are calculated for different values of $\Delta E$ (as marked) with $\rho(0)$ kept constant at $0.1n$. Black solid lines correspond to the oscillations shown in Fig.~\ref{de}(a)--(b) and \ref{de}(e)--(f). The red solid lines mark the separatrix trajectories at the critical values $\Delta E_c^{\pm}$ (oscillations corresponding to these trajectories are shown in Fig.~\ref{de}(c) and \ref{de}(f)). Black dashed lines correspond to the a.c. Josephson regime of oscillations (see Fig.~\ref{de}(d) and \ref{de}(h)), when the values of $\Delta E$ exceed the critical values $\Delta E_c^{\pm}$.

\section{Conclusion}\label{conclusion}

We have investigated the dynamical regimes of internal oscillations in a tuned two-component Bose condensate with interaction only in one of the components. Introducing the set of coupled temporal Gross-Pitaevskii and Schr\"{o}dinger equations, we described the evolution of population imbalance and relative quantum phase between the subsystems for different values of energy detunings, being guided by the physical system of exciton-polaritons in an optical microcavity. We show that, depending on the value of effective detuning, the modified harmonic or anharmonic (depending on the initial conditions) Rabi oscillations between the two particle states can be replaced by an analog of internal a.c. Josephson effect. Also, we predict that there is a defined value of the energy detuning close to zero yet depending on the initial population imbalance, at which the internal oscillations are completely suppressed. For polariton BEC, that would mean that the photon and exciton components, although still in resonance, stop exchanging particles, the population imbalance is constant and the relative phase is fixed at the value $\pi$, which means that all particles in the condensate are locked in the lower polariton state.

\section*{Acknowledgments}
This work is partially supported by Russian Foundation for Basic Research Grant No.~14-22-02091. The work of N.S.V. is partially supported by NRNU MEPhI ``Academic mobility'' program.

\end{document}